\documentclass[12pt,preprint]{aastex}

\newcommand{\logg} {\log g}
\newcommand{\halpha} {H$\alpha$}
\newcommand{\Te} {T_{\rm eff}}
\newcommand{\mv} {$M_V$}
\newcommand{\htwo} {H$_2$}
\newcommand{\msun} {$M_\odot$}
\newcommand{\vtan} {v_{\rm tan}}
\newcommand\kms{km~s$^{-1}$}
\newcommand{\nhe} {N({\rm He})/N({\rm H})}
\newcommand{\nh} {N({\rm H})/N({\rm He})}
\shortauthors{Bergeron et al.}
\shorttitle{High Velocity White Dwarfs}
\begin{document}

\title{On the Interpretation of High Velocity White Dwarfs as 
Members of the Galactic Halo}

\author{P. Bergeron\altaffilmark{1}, Mar\'\i a Teresa Ruiz\altaffilmark{2},
M.~Hamuy\altaffilmark{3}, S.~K. Leggett\altaffilmark{4}, M.~J.~Currie\altaffilmark{5},}
\author{C.-P. Lajoie\altaffilmark{1}, and P. Dufour\altaffilmark{1}}

\altaffiltext {1} {D\'epartement de Physique, Universit\'e de Montr\'eal, 
C.P.~6128, Succ.~Centre-Ville, Montr\'eal, Qu\'ebec, Canada, H3C 3J7; 
bergeron,lajoie,dufour@astro.umontreal.ca.}

\altaffiltext {2} {Departamento de Astronom\'\i a, Universidad de Chile, 
Casilla 36-D, Santiago, Chile; mtruiz@das.uchile.cl.}

\altaffiltext {3} {Las Campanas Observatory, Carnegie Observatories, 
Casilla 601, La Serena, Chile; mhamuy@lco.cl.}

\altaffiltext {4} {UKIRT, Joint Astronomy Centre, 660 North A'ohoku Place, 
Hilo, HI 96720; s.leggett@jach.hawaii.edu.}

\altaffiltext {5} {Starlink Project, Rutherford Appleton Laboratory, Chilton, 
Didcot, Oxon OX11 0QX, UK; mjc@star.rl.ac.uk.}

\begin{abstract}

A detailed analysis of 32 of the 38 halo white dwarf candidates
identified by Oppenheimer et al.~is presented, based on model
atmosphere fits to observed energy distributions built from optical
$BVRI$ and infrared $JHK$ CCD photometry. Effective temperatures and
atmospheric compositions are determined for all objects, as well as
masses and cooling ages when trigonometric parallax measurements are
available. This sample is combined with that of other halo white dwarf
candidates and disk white dwarfs to study the nature of these objects
in terms of reduced proper motion diagrams, tangential velocities, and
stellar ages. We reaffirm the conclusions of an earlier analysis based
on photographic magnitudes of the same sample that total stellar ages
must be derived in order to associate a white dwarf with the old halo
population, and that this can only be accomplished through precise
mass and distance determinations.

\end{abstract}


\section{Introduction}

White dwarf stars cool slowly enough that even the coolest and thus
oldest white dwarfs are still visible \citep[see][for a
review]{fon01}. The abrupt cutoff in the observed luminosity function
of white dwarfs has been used by
\citet{win87} to infer the age of the local Galactic disk.
A more recent determination by \citet{lrb98} using the 43 white dwarfs
in the original sample of \citet{ldm88} lead to an age estimate of
$8\pm1.5$ Gyr. White dwarf stars are also being used in globular
clusters to get an independent estimate of the cluster ages
\citep{hansen04}.

There has been a growing interest in identifying white dwarfs in the
old halo population of our Galaxy, primarily to determine whether
these old remnants could contribute significantly to the halo dark
matter. \citet[][hereafter OHDHS]{opp01a} claimed to have discovered
such a population by identifying 38 cool halo white dwarf candidates
in the SuperCOSMOS Sky Survey, with an inferred space density that
could account for 2\% of the halo dark matter. The various criticisms
that followed that study \citep[see, e.g.][]{reid01,hansen01,torres02}
were re-examined by \citet{salim04} who basically confirmed the
conclusions reached by OHDHS. Most of these studies looked at these
halo white dwarf candidates from the point of view of their kinematics.

\citet{ber03} analyzed the photographic magnitudes obtained by OHDHS
using model atmosphere fits to observed energy distributions following
the photometric method described at length in \citet[][hereafter BRL]{brl97} and
\citet[][hereafter BLR]{blr01}. The analysis suggested that most of
the white dwarfs in the OHDHS sample were probably too hot and too
young to be associated with the halo population of the Galaxy. In this
paper we present a similar analysis based on CCD photometry rather
than photographic magnitudes, and with the addition of near-infrared
photometry. Our photometric observations and theoretical framework
are described respectively in \S~2 and 3. The results of our analysis
in terms of reduced proper motion diagrams, tangential velocities, and
stellar ages are then presented in \S~4. Our conclusions follow in
\S~5.

\section{Photometric Observations}

Optical $BVRI$ CCD photometry has been secured for 30 white dwarfs
taken from the OHDHS sample during several runs in 2002 and 2003, at
Las Campanas (Carnegie Observatories) using the 1.3 m Warsaw
telescope and the 1 m Swope telescope. Photometric standards from
\citet{landolt83} were used for calibration. Our photometry is
reported in Table 1 together with the number of independent
observations ($N$) for each object. Uncertainties are approximately
5\% at $B$ and 3\% at $VRI$. For the two stars with $N=0$, we have
used the photographic magnitudes of OHDHS transformed into the
standard $V$ and $I$ magnitudes using the relations defined in
equations (1) to (5) of \citet{salim04}. Our optical photometry is
compared in Figure \ref{fg:f1} with that of
\citet[][their Table 2]{salim04}. The agreement between both
photometric sets is excellent. The largest discrepancy is for the $V$
magnitude of WD~0205$-$053\footnote{Here and in the following we use
for consistency the object names as defined in OHDHS. Note, however,
that when OHDHS assigned a WD number to an object, it was based on
2000 coordinates while it would have been more appropriate to use the
1950 coordinates following the rules of the white dwarf catalog of
\citet{mccook99}. The correct WD numbers are used in the WD column of
our Table 1.} for which Salim et al.~report a value 0.31 mag fainter
than our measurement, with an uncertainty of 0.145 mag.

The infrared $JHK$ photometry for 22 of the white dwarfs in Table 1
was obtained between 2002 September 23 and 26 using ClassicCam on
Magellan.  Four more stars were also observed at Las Campanas 
using the Dupont (100-inch) telescope and the WIRC camera, on
2003 December 14 and 15. One other was
observed on the UK Infrared Telescope (UKIRT) using the UFTI camera,
on 2004 June 26.  These observations were calibrated using either the
photometric standards of \citet{persson98} or \citet{hawarden01}.
The ClassicCam data were reduced using the software described in
\citet{currie04}. Infrared data for LHS 147 and LHS 542 are taken from
BLR, and for LHS 4033 from \citet{dahn04}. Out of the 32 objects
listed in Table 1, 22 have $JHK$ measurements, 8 only have $J$ and
$H$, while 2 have no infrared data.  Also reported in Table 1 are the
infrared photometric uncertainties (in parentheses) and the number of
independent observations.

\section{Theoretical Framework}

The model atmospheres used in this analysis are described at length in
\citet[][see also BRL and BLR]{bsw95} with the 
collision-induced opacities from molecular hydrogen updated from the
work of \citet{jorgensen} and \citet{borysow01}. These models are in
local thermodynamic equilibrium, they allow energy transport by
convection, and they can be calculated with arbitrary mixed hydrogen
and helium compositions.

Synthetic colors\footnote{These synthetic colors can be obtained at
http://www.astro.umontreal.ca/\~{ }bergeron/CoolingModels} are
obtained using the procedure outlined in
\citet{bwb95} but with the new Vega fluxes taken from \citet{bohlin04} 
and the Vega magnitudes from Table A1 of \citet{bessell98}. Similarly,
in order to compare the photometric observations with the model
atmosphere predictions, we convert (see also BRL) the optical and
infrared magnitudes $m$ into observed fluxes averaged over the
transmission function $S_m(\lambda)$ using the following equation

$$m= -2.5\log f_{\lambda}^m + c_m\ , \eqno (1)$$

\noindent where
$$f_{\lambda}^m= {{\int_0^\infty
f_{\lambda}S_m(\lambda)d\lambda}\over{\int_0^\infty
S_m(\lambda) d\lambda}}\eqno (2)$$

\noindent is the averaged observed flux received at Earth. 
The transmission functions $S_m(\lambda)$ are taken from
\citet{bessell90} for the $BVRI$ filters on the Johnson-Cousins
photometric system, and from
\citet{bessell88} for the $JHK$ filters on the Johnson-Glass
system. The constants $c_m$ for each passband using the new fluxes and
zero points for Vega are $c_B=-20.4761$, $c_V=-21.0798$,
$c_R=-21.6300$, $c_I=-22.3480$, $c_J=-23.7417$, $c_H=-24.8387$, and
$c_K=-25.9877$. These constants differ slightly from those used by BRL
and BLR, which were based on older Vega fluxes.  Note also that with
this new calibration, the +0.05 mag correction determined empirically
and applied by BRL to the $J$, $H$, and $K$ constants is not required
here (see \S~5.2.1 of BRL).

Since some of our observed magnitudes were obtained on the infrared
system defined by \citet{persson98}, we also calculated theoretical
colors using the filter passbands described in their Appendix, but
found negligible differences with the calculations using the
Johnson-Glass system. We thus rely only on the latter in our analysis.

\section{Results}

\subsection{Two-color Diagrams}

We first present in Figure \ref{fg:f2} the ($V$--$I$, $V$--$H$)
two-color diagram for 29 objects from Table 1. Spectroscopic
observations obtained from B.~R.~Oppenheimer (2004, private
communication) are used to discriminate between DA (i.e.~spectra
showing \halpha) and non-DA stars. Also shown are the theoretical
colors for our pure hydrogen, pure helium, and $\nh=10^{-5}$ model
atmospheres. The loops observed in this diagram for the models
containing hydrogen are the result of the presence of the
collision-induced opacity from molecular hydrogen that reduces the
flux significantly in the infrared. The effect occurs at higher
effective temperature in the models at $\nh=10^{-5}$ because despite
the fact that the abundance of hydrogen is greatly reduced, the higher
atmospheric pressure of these models increases the number of
collisions, which in turn increases the contribution of the \htwo-He
collision-induced opacity. Smaller or larger hydrogen abundances would
yield smaller infrared flux deficiencies according to the calculations
of \citet[][see their Fig.~5]{bl02}.

DA stars (filled circles) follow nicely the hydrogen sequence. Since
the pure hydrogen and pure helium sequences start at $\Te=12,000$~K,
the location of the hottest DA stars in Figure \ref{fg:f2}
already suggests that several stars in the OHDHS sample are very hot.
One non-DA star overlapping the DA stars, LHS 1447, is warm enough
to show \halpha\ according to its location in Figure
\ref{fg:f2}, a result that suggests it
probably has a helium-rich atmospheric composition. At lower effective
temperatures, $\Te<5000$~K, \halpha\ disappears altogether --- even in
pure hydrogen atmospheres --- because of the Boltzmann factor. Hence
we can no longer rely on the presence of \halpha\ to infer the
atmospheric composition of the white dwarf and fits to the energy
distribution must be used instead, as discussed in the BRL and BLR
analyses.  As can be seen from Figure
\ref{fg:f2}, the pure hydrogen and pure helium sequences cross
each other at
low temperatures, which makes the discrimination between both
atmospheric compositions a difficult task. This problem is less severe
when the entire energy distribution is used, however (see below).

Two objects are labeled in Figure \ref{fg:f2}: WD~2356$-$209,
further discussed in \S \ref{sbsc:wd2356}, is a cool white dwarf with
an odd spectrum according to OHDHS, with a strong absorption feature
near 6000 \AA\ that strongly affects the $V$ magnitude in Figure
\ref{fg:f2}. LHS 1402, further discussed in \S
\ref{sbsc:lhs1402}, is another extremely cool white dwarf candidate showing 
a very strong infrared flux deficiency similar to those observed in
LHS 3250 and SDSS 1337+00, or in the handful of candidates identified
by \citet{gates04}. As for LHS 3250 and SDSS 1337+00, the location of LHS 1402 in
the ($V$--$I$, $V$--$H$) two-color diagram suggests either an
extremely cool hydrogen-atmosphere white dwarf, or a much warmer star
with a helium-rich atmospheric composition.

\subsection{Energy Distributions}

To derive the atmospheric parameters for each star in our sample, we
rely on the technique developed by BRL, which we briefly
describe again here for completeness. To make use of all the
photometric measurements simultaneously, we convert the magnitudes
into observed fluxes using equation (1), and compare the resulting
energy distributions with those predicted from our model atmosphere
calculations. For each star, we obtain a set of seven (or less)
average fluxes $f_{\lambda}^m$ which can now be compared with the
model fluxes. These model fluxes are also averaged over the filter
bandpasses by substituting $f_{\lambda}$ in equation (2) for the
monochromatic Eddington flux $H_{\lambda}$. The average observed
fluxes $f_{\lambda}^m$ and model fluxes $H_{\lambda}^m$ --- which
depend on $\Te$, $\logg$, and $\nhe$ --- are related by the equation

$$f_{\lambda}^m= 4\pi~(R/D)^2~H_{\lambda}^m\ , \eqno (3)$$

\medskip
\noindent where $R/D$ is the ratio of the radius
of the star to its distance from Earth. Our fitting procedure relies
on the nonlinear least-squares method of Levenberg-Marquardt, which is
based on a steepest descent method.  The value of $\chi ^2$ is taken
as the sum over all bandpasses of the difference between both sides of
equation (3), properly weighted by the corresponding observational
uncertainties. In our fitting procedure, we consider only $\Te$ and
the solid angle free parameters.

As discussed by BRL, the energy distributions are not sensitive enough
to surface gravity to constrain the value of $\logg$, and thus for
white dwarfs with no parallax measurement, we simply assume
$\logg=8.0$.  For stars with known trigonometric parallax
measurements, we start with $\logg=8.0$ and determine $\Te$ and
$(R/D)^2$, which combined with the distance $D$ obtained from the
trigonometric parallax measurement yields directly the radius of the
star $R$.  The radius is then converted into mass using the cooling
sequences described in BLR with thin and thick hydrogen layers, which
are based on the calculations of \citet{fon01}. In general, the
$\logg$ value obtained from the inferred mass and radius will be
different from our initial guess of $\logg=8.0$, and the fitting
procedure is thus repeated until an internal consistency in $\logg$ is
reached.

Only three objects in our sample have trigonometric parallax
measurements, LHS 147 ($14.0\pm9.2$ mas), LHS
542 ($32.2\pm3.7$ mas), and LHS 4033 ($33.9\pm0.6$ mas). The value for
LHS 4033 is taken from
\citet{dahn04}, while the values for the other stars correspond to much 
older measurements with corresponding larger uncertainties. 
We note that the uncertainty for LHS 147 is as much
as 65 \%. More modern unpublished measurements obtained by the US
Naval Observatories indicate that the above values have not changed
significantly, but the uncertainties have been greatly reduced
(H.~C.~Harris, 2004, private communication).

Sample fits for four objects in our sample are displayed in Figure
\ref{fg:f3}. The left panels compare our best solutions with 
pure hydrogen and pure helium atmospheric compositions, while the
right panels show the observed spectra obtained by OHDHS near the
\halpha\ region together with the model spectrum calculated from the pure 
hydrogen solution. Together, the left and right panels can be used to
determine the atmospheric composition and effective temperature of
each star. We explore here only pure hydrogen and pure helium
atmospheric compositions; limits on traces of hydrogen or helium in
cool white dwarf atmospheres have been discussed in BRL.  Note that
because the stars in the OHDHS sample are much fainter than those
studied in BRL and BLR, the quality of the fits to the energy
distributions are not as good.

WD~0100$-$645 represents a good example of
a pure hydrogen atmosphere white dwarf. Even though the hydrogen
($\chi^2=4.0$) and helium ($\chi^2=5.3$) fits do not differ much, the
presence of the \halpha\ feature clearly favors the hydrogen solution.
The inferred effective temperature is also consistent with the
observed \halpha\ line profile. We note, however, that the predicted
flux at $I$ is outside the $1~\sigma$ observational uncertainty with
the hydrogen fit, suggesting that the measured flux at $I$ may be in
error.  This emphasizes the importance of using the complete $BVRI$
and $JHK$ energy distributions to study these faint objects instead of
using color-color diagrams, which tend to accentuate these errors in
the photometric measurements.

The second object in Figure \ref{fg:f3},
LHS 1447, is a good example of a pure helium atmosphere white
dwarf. In this case, the $\chi^2$ value of the helium fit (5.9) is
much smaller than that of the hydrogen fit (23.7). In particular, the
hydrogen model fails to reproduce the flux in the $H$ bandpass within
the uncertainties. This is related to the fact that the H$^-$ opacity,
which dominates in this temperature range, has a minimum at 1.6
\micron\ (bound-free threshold), producing a local maximum in the
energy distribution of hydrogen models. Furthermore, the predicted
\halpha\ line profile assuming a pure hydrogen atmosphere for LHS 1447
clearly rules out this solution.

The other two objects, F351$-$50 and
WD~0227$-$444, are too cool to show \halpha, even if we assume a pure
hydrogen composition. Hence we must rely solely on the fits to the
energy distributions. F351$-$50 represents an excellent example of a
cool, pure hydrogen atmosphere white dwarf. The differences between
the hydrogen and helium solutions are extreme in this case
($\chi^2=3.6$ for the hydrogen fit as opposed to $\sim 150$ for the
helium fit). Our pure hydrogen fit, however, fails to reproduce the
observed flux at $B$ within the uncertainties, and also at $V$ to a
lesser extent. This discrepancy has been explained by BRL in terms of
a missing opacity source in the ultraviolet of the pure hydrogen
models, most likely due to a pseudo-continuum opacity originating from
the Lyman edge (see \S\ 5.2.2 of BRL for a complete description),
although this explanation has been challenged by
\citet{wolff02}. Note also that the failure of the pure hydrogen
models to match the observed fluxes in this particular region of the
energy distribution is most likely at the origin of the peculiar
solution obtained for F351$-$50 by
\citet{opp01b} -- $\Te=2844$ ~K, $\logg=6.5$, and $\nhe=0$ (see 
their Fig.~9) -- based solely on a spectrum covering the region
between 0.4 and 1 $\mu$m.

Finally, WD~0227$-$444, shown at the
bottom of Figure \ref{fg:f3}, represents a good example of a cool
white dwarf with a pure helium atmosphere ($\chi^2=8.9$ as opposed to
25.1 for the hydrogen fit). In this case, only the observed flux at
$H$ is not matched by the helium model, within the uncertainties. In
contrast, six out the seven bands used in our fitting procedure are
not matched properly by the hydrogen model.

The atmospheric parameters $\Te$, $\logg$, and atmospheric composition
(H or He) for the 32 objects listed in Table 1 are given in Table 2
together with the calculated stellar mass, absolute visual magnitude,
luminosity, distance, and white dwarf cooling age. The latter is
obtained from the theoretical cooling sequences described above. A
value of $\logg=8.0$ was assumed for all stars expect where noted;
photometric distances are given for these objects. Three objects in
the OHDHS sample stood out in our analysis, WD~2356$-$209 and LHS 1402
labeled in Figure \ref{fg:f2}, which had to be analyzed in
greater detail. We discuss them in the next two sections. The third
object is the extremely massive DA white dwarf LHS 4033 analyzed in
detail by \citet{dahn04}.

\subsubsection{WD~2356$-$209}\label{sbsc:wd2356}

WD~2356$-$209 whose spectrum is
shown in Figure 2 of OHDHS and reproduced here in Figure \ref{fg:f4},
exhibits a strong absorption feature near 6000 \AA, which has been
interpreted by \citet{salim04} as possibly originating from an
extremely broad Na I doublet. A similar object has also been reported
by \citet[][see SDSS J1330+6435 in their Fig.~10]{harris03}.
Indeed, our modeling of the Na I D
doublet in a helium-rich atmosphere matches the observed broadband
energy distribution and the observed spectrum quite well (see
Fig.~\ref{fg:f4}). However, it was not possible to constrain
effectively the sodium abundance in this object since variations in
the sodium abundance could be compensated by changing the effective
temperature ($\pm 200$~K for $\pm1$ dex in sodium abundances) with
very little changes in the predicted spectrum in the wavelength range
used here. Large differences are predicted shortward of 5000 \AA,
however, and high signal-to-noise spectroscopy in this region should
help constrain better the abundances of sodium and other heavy
elements in the atmosphere of WD~2356$-$209, as well as its effective
temperature. Indeed, all the spectral features predicted in this
region of the spectrum are sodium lines. For the moment, we adopt a
solution with a sodium abundance close to the solar abundance, $N({\rm
Na})/N({\rm He})=10^{-5}$ and $\Te=4790$~K, which produces enough
blanketing in the optical to deplete the flux near the $B$
filter. This abundance may seem extreme but nearly solar abundances of
iron and magnesium have also been measured in the cool and massive DAZ
star GD 362 \citep{gianninas04}. 

\subsubsection{LHS 1402}\label{sbsc:lhs1402}

LHS 1402 whose spectrum is shown in Figure 2 of OHDHS and reproduced
here in Figure \ref{fg:f5}, exhibits a strong infrared flux
deficiency similar to those observed in LHS 3250 and SDSS 1337$+$00,
and in the ultracool white dwarf candidates reported by \citet[][their
Fig.~2]{gates04}.  The detailed photometric and model atmosphere
analysis of the first two objects by \citet{bl02} has revealed that
the infrared flux deficiency, steep optical spectrum, and luminosity
(known only for LHS 3250)
could be explained better in terms of an extremely helium-rich
atmospheric composition rather than a pure hydrogen composition. In
the latter case, the infrared flux deficiency is the result of
collision-induced absorptions by molecular hydrogen, a mechanism that
becomes important only at very low temperatures when the collisions
responsible for the absorption are between hydrogen molecules
only. However, in a helium-rich environment, characterized by higher
atmospheric pressures, collisions also occur with neutral helium. The
overall result is that it is possible to reproduce the same infrared
flux deficiency but at much higher effective temperatures and
luminosities, in better agreement with the observations. Furthermore,
the broad absorption feature near 0.8 $\mu$m predicted by the pure
hydrogen models is simply not observed \citep[see Fig.~7 of][]{bl02}.

This situation is similar for LHS 1402, as shown in Figure
\ref{fg:f5}, where we contrast our best solutions for a pure 
hydrogen composition and a mixed hydrogen and helium composition.  As
for LHS 3250 and SDSS 1337+00, both solutions fail to reproduce
adequately the peak of the energy distribution, although the
helium-rich solution exhibits a broader peak, not as high as that of
the hydrogen solution, in closer agreement with the observations. The
reasons for this discrepancy is still being investigated by us and
others \citep[see, e.g.,][]{kowalski04}.  As discussed above, the dip
near 0.8 $\mu$m predicted by the pure hydrogen solution is simply not
observed, a result that suggests that the atmosphere of LHS 1402 is
indeed helium rich. Since the effective temperatures inferred from both
solutions differ by over 1000 K, a measurement of the trigonometric
parallax and thus of the absolute visual magnitude should help
discriminate between our two solutions, as was done for LHS 3250 by
\citet[][see their Fig.~8]{bl02}. In the following, we adopt the
atmospheric parameters from our solution with $\log\nh=-4.5$ shown in
Figure \ref{fg:f5}. Note that the white dwarf cooling age
obtained from the helium-rich solution, 9.86 Gyr given in Table 2, is
significantly shorter than that derived from the hydrogen solution,
12.5 Gyr.

\subsection{Reduced Proper Motion Diagram}

One very important tool that is commonly used in identifying halo
white dwarf candidates is the reduced proper motion diagram.  The
reduced proper motion combines an observed magnitude with the proper
motion measurement to yield some estimate of the absolute magnitude of
the star \citep[see][]{knox99}.  OHDHS and \citet{ber03} relied on a reduced
proper motion defined as $H_{\rm R}=R_{\rm 59F}+5\log\mu+5$, where
$R_{\rm 59F}$ is the photographic magnitude, and $\mu$ is the proper
motion measured in arc seconds per year, and those values were plotted
against the photographic color index $B_{\rm J}$--$R_{\rm 59F}$.
Stars that are relatively blue and with large values of $H_{\rm R}$ in
this diagram are viewed as good halo white dwarf candidates since old,
and thus cool, white dwarfs have low luminosities and turn blue below
$\sim 3500$~K. Moreover, white dwarfs belonging to different kinematic
populations of the Galaxy will be well separated in this diagram.

Our improved reduced proper motion diagram using CCD photometric
measurements is displayed in Figure \ref{fg:f6} where the reduced proper
motion $H_{\rm V}=V+5\log\mu+5$ is plotted against the $V$--$I$ color
index. The open circles represent the data taken from the BRL and BLR
samples, while the filled symbols correspond to the data taken from Table
1. The objects labeled in the Figure represent five of the six halo
white dwarf candidates identified by \citet{ldm89} on the basis of
their large tangential velocities. Two of these stars, LHS 147 and LHS
542, are in common with the OHDHS sample. The leftmost and rightmost
objects in Figure \ref{fg:f6} correspond to LHS 1402 and WD
2356$-$209, respectively, while the two stars at the bottom are the
hydrogen-rich white dwarfs F351$-$50 and WD~0351$-$564, two of the
coolest objects in Table 2.

Also differentiated in Figure \ref{fg:f6} are the stars above and
below $\Te=5000$~K. With the exception of LHS 1420, all stars below
5000~K (filled circles) overlap with the (extended) sequence defined
by the disk sample of BRL and BLR. All stars to the left of this
sequence have temperatures above 5000~K (filled diamonds).  With the
exception of LHS 542 at $\Te=4740$~K, all halo white dwarf
candidates from \citet{ldm89} also have temperatures in excess of
5000~K. It thus appears that most objects identified in such reduced
proper motion diagrams are not cool and old white dwarfs, but instead
relatively hot white dwarfs with large proper motions, and presumably
large tangential velocities (see next section). Even LHS 1402 appears
relatively luminous for its blue $V$--$I$ color index, most likely
because the infrared flux deficiency that characterizes its energy
distribution is the result of \htwo-He collision-induced absorptions
in a warm, helium-rich atmosphere, as opposed to \htwo-\htwo\
collision-induced absorptions in an extremely cool, hydrogen-rich
atmosphere. Cool and old pure hydrogen atmosphere white dwarfs would
reside at much larger values of the reduced proper motion.

\subsection{Tangential Velocities}

The kinematic analysis of \citet{salim04} in the $U-V$ plane
velocities relies heavily on the distance estimates (see their Fig.~5
for instance). We compare in Figure \ref{fg:f7} our
own distance estimates with those given in Table 4 of Salim et
al. Surprisingly, despite the much cruder approach used by Salim et
al.~to estimate individual distances, the results agree extremely
well. The only noticeable exception is for LHS 4033 whose
trigonometric parallax measurement implies a distance much closer and
a mass much larger ($M=1.34$ \msun) than that obtained under the
assumption of $\logg=8.0$.  We thus conclude that the kinematic
analysis of Salim et al.~will not be affected by our results, and will
thus not be repeated here. We simply reaffirm the conclusions of Salim
et al.~that the kinematics in the $U$- and $V$-components of the
velocity plane of the OHDHS sample are consistent with a mixed of
thick-disk and halo white dwarfs.

We look instead at the distribution of tangential velocities $\vtan$
with absolute visual magnitudes \mv\ for the OHDHS sample, displayed
in Figure \ref{fg:f8}. The tangential velocities are calculated
using the proper motions provided in Table 4 of \citet{salim04} and
the distances taken here from Table 2. Also shown are the results for
the trigonometric parallax sample of BLR, which includes three of the
five halo white dwarf candidates from \citet{ldm89} labeled in Figure
\ref{fg:f8}; LHS 282 and LHS 291 have trigonometric parallax
measurements that are too uncertain to derive meaningful
distances. LHS 147 and LHS 542 in common between the BLR and OHDHS
samples demonstrate the repeatability of our atmospheric parameter
measurements. The object at the very bottom is the massive white dwarf
ESO 439$-$26 with an estimated mass of $\sim 1.2$ \msun, an effective
temperature of 4500~K, and an absolute visual magnitude of \mv=17.46.

It is already clear that the tangential velocities of the OHDHS sample
differ quite markedly from those of the disk stars. And indeed some
objects have tangential velocities well in excess of 200 \kms. The
most extreme case is for WD 0135$-$039 with a value of $\vtan=430$
\kms. This object is not particularly cool, however, with a temperature 
of 7470~K. Again we note that none of these objects are particularly
cool (the right axis indicates the temperature scale for 0.6 \msun\
white dwarf models). Even the coolest object in our analysis, LHS 1402, 
has the smallest tangential velocities of all ($\vtan=60$ \kms). The
objects with the largest tangential velocities even have tendencies to
be located hotter than 7000~K, the three exceptions being LHS 542, WD
0351$-$564, and F351$-$50.

\subsection{Stellar Ages}

Insight into the nature of the halo
white dwarf candidates identified by OHDHS may be gained by estimating
their total stellar ages. Halo white dwarfs should have total ages
well in excess of 10 Gyr. We show in Figure \ref{fg:f9} the location
in a mass versus effective temperature diagram of all white dwarfs
taken from the parallax sample of BLR (open symbols) and the OHDHS
sample (filled symbols). Various symbols explained in the legend are
used to differentiate ranges of tangential velocities. White dwarfs
from the OHDHS sample with no trigonometric parallax measurements and
for which it is not possible to determine the mass are shown at the
bottom of the Figure.

Mass uncertainties are also indicated for all white dwarfs in the
OHDHS sample with measured parallaxes (LHS 147, LHS 542, and LHS 4033)
and for the white dwarfs in the BLR sample with $\vtan>200$ \kms\ (LHS
56, LHS 147, and LHS 542; the last two objects are in common with the
OHDHS sample and they have identical error bars). Unfortunately, these
mass uncertainties are fairly large, with the exception of LHS 4033 at
$M\sim 1.3$ \msun, which corresponds to a modern parallax measurement
\citep{dahn04}. As discussed in \S~4.2, however, both LHS 147 and LHS 542 
have been measured with comparable accuracy, and the parallax values
have not changed significantly from those used in Figure \ref{fg:f9}
(H.~C.~Harris, 2004, private communication).

Also superposed on this plot are the theoretical isochrones from the
white dwarf cooling sequences discussed above with C/O-cores, $q({\rm
He})\equiv M_{\rm He}/M_{\star}=10^{-2}$, and $q({\rm H})=10^{-4}$.
The solid lines represent the white dwarf cooling ages only. These
parabola-shaped isochrones are the result of the onset of
crystallization occuring first in the higher mass models, reducing the
cooling timescales considerably. With decreasing effective
temperature, crystallization gradually occurs in lower mass models,
and the turning point of these parabola moves slowly towards lower
masses. Since {\it total stellar ages} and not white dwarf cooling
ages are the crucial aspect we want to investigate here, we must take
into account the time spent on the main sequence. To do so, we follow
the procedure outlined in \citet{wood92} and we add to the white dwarf
cooling age the main sequence lifetime $t_{\rm MS}$ calculated as
$t_{\rm MS}=10(M_{\rm MS}/M_\odot)^{-2.5}$ Gyr where $M_{\rm MS}$ is
the mass on the main sequence of the white dwarf progenitor. The
latter is obtained from the initial-final mass relation for
white dwarfs, a relation that is not particularly well determined,
especially at low mass \citep[see][for a review]{weidemann00}. 
Here we use the parameterization used by \citet{wood92}

$$M_{\rm WD}= A_{\rm IF}\exp(B_{\rm IF}M_{\rm MS})\ , \eqno (4)$$

\medskip
\noindent where $M_{\rm WD}$ is the mass of the white dwarf, and 
$A_{\rm IF}$ and $B_{\rm IF}$ represent constants that need to be
determined empirically. \citet{wood92} used the spectroscopic mass
distribution of DA white dwarfs obtained by \citet{bsl} and derived
$A_{\rm IF}=0.4$ and $B_{\rm IF}=0.125$. The mass distribution of
Bergeron et al.~relied on thin hydrogen layer models, while thick
hydrogen models yield larger masses \citep{brb95}. Since the weight of
evidence now is that most DA white dwarfs have thick outer hydrogen
layers, we redetermined the constants in equation (4) by using the
mass distribution obtained for the 348 DA stars from the Palomar-Green
survey sample \citep{lbh05}, which is based on the thick hydrogen
evolutionary models of \cite{wood95}. We obtain the following
constants, $A_{\rm IF}=0.45$ and $B_{\rm IF}=0.144$. Thus a main
sequence star with a 12 Gyr lifetime -- corresponding to a mass of
0.93 \msun\ -- would produce a 0.51 \msun\ white dwarf remnant.

The above empirical initial-final mass
relation has been derived using white dwarfs from the thin disk, while
the relation for halo white dwarfs --- which is
even more poorly known --- is probably similar to that of globular
clusters. As discussed by
\citet{renzini96}, the mass of the white dwarfs currently being formed
in globular clusters can be constrained by the luminosities of the red
giant branch tip, the horizontal branch, the AGB termination, and the
post-AGB stars, all of which are sensitive to the mass of the hydrogen
exhausted core. All observations point to values between $M_{\rm
WD}=0.51$ and 0.55 \msun, virtually independent of metallicity
\citep{renzini88}. Hence it is reasonable to assume that white dwarfs
currently being formed in the halo should have masses in the same
range. The empirical initial-final mass relation we derived above is certainly
consistent with these results, although it should be considered a
good approximation at best, and white dwarfs currently being formed in the
halo could still be as massive as 0.55 \msun.

The isochrones representing the white dwarf cooling ages plus main
sequence ages using the initial-final mass relation described above are reproduced in
Figure \ref{fg:f9}. It is clear that the total age of a white dwarf is
strongly mass-dependent, a result which stresses the importance of
determining reliable masses through precise trigonometric parallax
measurements. For instance, all white dwarfs with masses below
$M\lesssim0.5$ \msun\ cannot have been formed within the lifetime of
the Galaxy, and they must be the result of common envelope
evolution. Alternatively, these could be unresolved degenerate
binaries, and their overluminosity would be wrongly interpreted here
as single white dwarfs with large radii and low masses (see BRL and
BLR for further discussion).  Also, the results of Figure \ref{fg:f9}
illustrate how a 12 Gyr old white dwarf, say, could be found at any
effective temperature, as long as its mass is precisely on the
horizontal part of the isochrones near $\sim0.5$ \msun, implying that
is has recently (a few Gyr) evolved from a main sequence star slightly
below $\sim 1$ \msun\ (see discussion above).

Only three white dwarfs from the OHDHS sample have trigonometric
parallax measurements. One of them is the extremely massive white
dwarf LHS 4033 \citep{dahn04} seen in the upper left corner of Figure
\ref{fg:f9}. So not only this star does not have the proper 
kinematics to be associated with the halo population, but it is also
much too young ($\tau<2$ Gyr). The other two objects, LHS 147 and LHS
542, have more normal masses of $M=0.64$ and 0.67 \msun,
respectively.  Taken at face value, they both appear too young to be
associated with the halo population, despite their halo
kinematics. However, when the mass uncertainties are taken into
account, their total stellar ages could be made consistent with the
age of the halo. This stresses the importance of reducing the size of
the parallax measurements through the use of modern CCD techniques,
such as those currently being obtained at the USNO.

If the values of the trigonometric parallax measurements for LHS 147
and LHS 542 are confirmed, these two white dwarfs could indeed be very
young according to our results. This conclusion seems to be
independent of the particular choice of the initial-final mass
relation adopted here since both stars have inferred masses nearly 0.1
\msun\ above the upper limit of 0.55 \msun\ for the white dwarfs
currently being formed in globular clusters, and presumably in the
galactic halo as well.

\section{Conclusions}

In this paper, we have demonstrated the
importance of determining {\it total} stellar ages in order to
associate any white dwarf with a given population. This can only be
accomplished through a precise mass determination, which for cool
white dwarfs require accurate trigonometric parallax
measurements. Even though it is not possible to conclude at this stage
that any white dwarf in the OHDHS sample is too young to belong to the
halo population, with the glaring exception of LHS 4033, modern
parallax measurements for at least two white dwarfs, LHS 147 and LHS
542, seem to indicate that young white dwarfs with halo kinematics do
exist. The possibility that that young high velocity white dwarfs, most
likely associated with the young disk, might exist is intriguing.
\citet{ber03} summarized some physical mechanisms proposed in the 
literature that could produce these young high-velocity white dwarfs.
These include remnants of donor stars from close mass-transfer
binaries that produced type Ia supernovae via the single degenerate
channel \citep{hansen02}, or other alternative mechanisms 
by which stars can be ejected from the thin disk into the galactic
halo with the required high velocities.

The other white dwarf stars in the OHDHS sample are fairly warm, and
the only way they could be associated with the halo population is to
have stellar masses near $\sim 0.51$ \msun, in which case they can indeed be
very old. Trigonometric parallaxes will hopefully become available for
all stars from this sample in the near future.  The two most
likely halo candidates in the OHDHS sample are F351$-$50 and
WD~0351$-$564 (the two objects at the bottom of Fig.~\ref{fg:f6} and
also labeled in Fig.~\ref{fg:f8}). They correspond to the two coolest
objects in Figure \ref{fg:f9} with $\vtan>200$ \kms\ (the two
rightmost filled circles at the bottom of the figure). Masses below
0.6 \msun\ would yield total stellar ages above 11 Gyr.

Based on the results of our analysis, we feel that any determination
of the space density of white dwarfs in the halo or even in the thick
disk based solely on a kinematic analysis is basically flawed, and one
must combine such analyses with a precise determination of total
stellar ages, which implies in turn that distance estimates must also
be obtained \citep[see, e.g.][]{pauli05}. Similarly, analyses based on
reduced proper motion diagrams are likely to reveal more of these
young high-velocity white dwarfs rather than the long sought old white
dwarf halo population.

\acknowledgements We thank B.~R.~Oppenheimer for providing us with his 
spectroscopic observations. This work was supported in part by the
NSERC Canada and by the Fund FQRNT (Qu\'ebec). MTR received partial
support from Fondecyt (1010404) and FONDAP (15010003).  Support for
this work was also provided to MH by NASA through Hubble Fellowship
grant HST-HF-01139.01A awarded bt the Space Telescope Science
Institute, which is operated by the Association of Universities for
Research in Astronomy, Inc., for NASA, under contract NAS 5-26555. The
United Kingdom Infrared Telescope is operated by the Joint Astronomy
Centre on behalf of the U.K.~Particle Physics and Astronomy Research
Council.

\clearpage
\clearpage
\begin{deluxetable}{p{1.3cm}lllllccccc}
\tabletypesize{\scriptsize}
\tablecolumns{11}
\tablewidth{0pt}
\tablecaption{Optical\tablenotemark{a} and Infrared Photometric Measurements}
\tablehead{
\colhead{WD\tablenotemark{b}} &
\colhead{Name} &
\colhead{$B$} &
\colhead{$V$} &
\colhead{$R$} &
\colhead{$I$} &
\colhead{$N$} &
\colhead{$J$} &
\colhead{$H$} &
\colhead{$K$} &
\colhead{$N$}}
\startdata
0011$-$399&J0014$-$3937  &19.28    &18.19    &17.57    &17.07    &2&16.43 (0.03)   &16.23 (0.02)   &16.17 (0.03)   &1\\
0041$-$286&WD 0044$-$284 &21.02    &19.87    &19.21    &18.70    &1&18.15 (0.03)   &17.99 (0.04)   &\nodata        &1\\
0042$-$064&WD 0045$-$061 &19.19    &18.26    &17.71    &17.22    &2&16.83 (0.02)   &16.59 (0.02)   &16.54 (0.04)   &1\\
0042$-$337&F351$-$50     &20.54    &19.01    &18.31    &17.67    &2&17.07 (0.02)   &17.04 (0.03)   &17.05 (0.05)   &1\\
0058$-$647&WD 0100$-$645 &17.77    &17.37    &17.14    &16.78    &1&16.57 (0.06)   &16.40 (0.06)   &16.34 (0.06)   &1\\
\\
0059$-$008&LP 586$-$51   &18.40    &18.18    &18.18    &18.07    &1&18.27 (0.10)   &18.30 (0.10)   &\nodata        &1\\
0115$-$270&WD 0117$-$268 &20.04    &19.04    &18.47    &18.02    &2&17.47 (0.02)   &17.20 (0.03)   &17.23 (0.10)   &1\\
0120$-$280&WD 0123$-$278 &20.93    &19.96    &19.41    &18.90    &1&18.29 (0.04)   &18.09 (0.04)   &18.05 (0.11)   &1\\
0133$-$042&WD 0135$-$039 &20.01    &19.68    &19.46    &19.24    &1&\nodata        &\nodata        &\nodata        &0\\
0133$-$548&WD 0135$-$546 &19.44    &18.37    &17.79    &17.29    &1&16.67 (0.02)   &16.44 (0.02)   &16.40 (0.04)   &1\\
\\
0136$-$340&LHS 1274      &17.59    &17.18    &16.88    &16.65    &1&16.44 (0.06)   &16.25 (0.06)   &16.12 (0.06)   &1\\
0145$-$174&LHS 147       &17.97    &17.62    &17.38    &17.16    &1&17.00 (0.05)   &16.85 (0.05)   &16.86 (0.05)   &2\\
0151$-$016&WD 0153$-$014 &18.90    &18.69    &18.57    &18.51    &1&18.29 (0.06)   &18.29 (0.06)   &\nodata        &1\\
0202$-$055&WD 0205$-$053 &19.91    &18.59    &17.85    &17.17    &2&16.56 (0.03)   &16.53 (0.03)   &16.46 (0.04)   &1\\
0212$-$420&WD 0214$-$419 &20.80    &19.81    &19.33    &18.64    &1&18.28 (0.04)   &18.03 (0.05)   &\nodata        &1\\
\\
0222$-$291&LHS 1402      &18.73    &18.05    &18.06    &18.49    &2&19.09 (0.05)   &19.43 (0.10)   &\nodata        &1\\
0225$-$446&WD 0227$-$444 &20.64    &19.53    &18.98    &18.41    &1&17.93 (0.04)   &17.78 (0.04)   &17.59 (0.07)   &1\\
0246$-$302&LHS 1447      &18.94    &18.50    &18.14    &17.90    &1&17.68 (0.06)   &17.65 (0.06)   &17.62 (0.10)   &1\\
0304$-$074&LP 651$-$74   &18.00    &17.35    &16.98    &16.62    &3&\nodata        &\nodata        &\nodata        &0\\
0338$-$331&WD 0340$-$330 &21.07    &19.76    &19.19    &18.65    &2&17.88 (0.05)   &17.71 (0.04)   &17.62 (0.06)   &1\\
\\
0343$-$363&WD 0345$-$362 &21.26    &20.23    &19.47    &18.94    &2&18.24 (0.05)   &18.10 (0.04)   &18.29 (0.10)   &1\\
0350$-$566&WD 0351$-$564 &22.11    &20.56    &19.72    &18.89    &1&18.44 (0.05)   &18.47 (0.06)   &\nodata        &1\\
2211$-$392&WD 2214$-$390 &16.41    &15.92    &15.59    &15.26    &1&14.92 (0.02)   &14.66 (0.02)   &14.65 (0.04)   &1\\
2239$-$199&WD 2242$-$197 &20.65    &19.74    &19.24    &18.87    &1&18.35 (0.03)   &18.08 (0.05)   &\nodata        &1\\
2256$-$467&WD 2259$-$465 &20.60    &19.56    &18.96    &18.48    &2&17.96 (0.03)   &17.83 (0.03)   &17.63 (0.08)   &1\\
\\
2316$-$064&LHS 542       &19.23    &18.15    &17.53    &16.99    &1&16.38 (0.05)   &16.12 (0.05)   &16.08 (0.05)   &2\\
2321$-$597&WD 2324$-$595 &16.98    &16.79    &16.77    &16.81    &1&16.84 (0.02)   &16.92 (0.03)   &16.97 (0.05)   &1\\
2343$-$481&WD 2346$-$478 &\nodata  &17.95    &\nodata  &17.11    &0&16.43 (0.02)   &16.17 (0.02)   &16.06 (0.03)   &1\\
2346$-$550&WD 2348$-$548 &19.70    &18.88    &18.41    &17.99    &2&17.45 (0.03)   &17.17 (0.03)   &17.16 (0.06)   &1\\
2349$-$031&LHS 4033      &17.17    &16.98    &\nodata  &16.91    &1&16.97 (0.05)   &16.92 (0.05)   &17.02 (0.05)   &1\\
\\
2352$-$326&LHS 4042      &\nodata  &17.41    &\nodata  &17.23    &0&17.05 (0.02)   &17.12 (0.03)   &17.96 (0.10)   &1\\
2354$-$211&WD 2356$-$209 &21.24    &21.03    &19.92    &18.78    &2&18.33 (0.04)   &18.28 (0.06)   &\nodata        &1\\
\enddata
\tablenotetext{a}{Optical photometric uncertainties are 5\% at $B$ and 3\% at $VRI$.}
\tablenotetext{b}{The WD numbers are based on 1950 coordinates while those defined by OHDHS are based on 2000 coordinates.}
\end{deluxetable}


\clearpage
\clearpage
\hoffset 0.0truein
\begin{deluxetable}{p{1.3cm}lrccclrclc}
\tabletypesize{\scriptsize}
\tablecolumns{12}
\tablewidth{0pt}
\tablecaption{Atmospheric Parameters of Halo White Dwarf Candidates}
\tablehead{
\colhead{} &
\colhead{} &
\colhead{} &
\colhead{} &
\colhead{} &
\colhead{} &
\colhead{} &
\colhead{$D$} &
\colhead{$v_{\rm tan}$} &
\colhead{Age$^a$} &
\colhead{}\\
\colhead{WD} &
\colhead{Name} &
\colhead{$\Te$ (K)} &
\colhead{log $g^b$} &
\colhead{Comp} &
\colhead{$M/$\msun} &
\colhead{$M_V$} &
\colhead{(pc)} &
\colhead{(km~s$^{-1}$)} &
\colhead{(Gyr)} &
\colhead{Notes}}
\startdata
0011$-$399&J0014-3937    & 4340 ( 70)                        &8.00           &H &0.58           &15.86          & 29&104&8.08           &                    \\
0041$-$286&WD 0044$-$284 & 4770 ( 50)                        &8.00           &He&0.57           &15.44          & 77&134&6.57           &                    \\
0042$-$064&WD 0045$-$061 & 5100 ( 50)                        &8.00           &He&0.57           &15.01          & 44&144&5.66           &                    \\
0042$-$337&F351-50       & 4100 ( 60)                        &8.00           &H &0.58           &16.22          & 36&408&8.76           &                    \\
0058$-$647&WD 0100$-$645 & 6900 (170)                        &8.00           &H &0.59           &13.64          & 55&145&1.57           &1                   \\
\\
0059$-$008&LP 586-51     &10210 (600)                        &8.00           &H &0.60           &12.10          &164&282&0.57           &1                   \\
0115$-$270&WD 0117$-$268 & 4920 ( 50)                        &8.00           &He&0.57           &15.23          & 58&131&6.21           &                    \\
0120$-$280&WD 0123$-$278 & 4880 ( 50)                        &8.00           &He&0.57           &15.27          & 86&149&6.29           &                    \\
0133$-$042&WD 0135$-$039 & 7470 (350)                        &8.00           &H &0.59           &13.32          &186&434&1.29           &1                   \\
0133$-$548&WD 0135$-$546 & 4800 ( 40)                        &8.00           &He&0.57           &15.37          & 39&125&6.48           &                    \\
\\
0136$-$340&LHS 1274      & 7000 (180)                        &8.00           &H &0.59           &13.59          & 52&143&1.52           &1                   \\
0145$-$174&LHS 147       & 7640 (180)                        &8.07           &H &0.64           &13.35          & 71&376&1.34           &1, 2                \\
0151$-$016&WD 0153$-$014 & 9000 (310)                        &8.00           &H &0.60           &12.59          &166&317&0.79           &1                   \\
0202$-$055&WD 0205$-$053 & 4170 ( 60)                        &8.00           &H &0.58           &16.19          & 30&147&8.56           &                    \\
0212$-$420&WD 0214$-$419 & 4910 ( 60)                        &8.00           &He&0.57           &15.22          & 82&130&6.21           &                    \\
\\
0222$-$291&LHS 1402      & 3240 ( 70)                        &8.00           &He&0.57           &15.98          & 25& 60&9.86           &3                   \\
0225$-$446&WD 0227$-$444 & 4880 ( 50)                        &8.00           &He&0.57           &15.26          & 71&117&6.30           &                    \\
0246$-$302&LHS 1447      & 6550 (170)                        &8.00           &He&0.57           &13.82          & 86&221&1.95           &                    \\
0304$-$074&LP 651-74     & 5750 (190)                        &8.00           &H &0.59           &14.41          & 38& 87&2.56           &1                   \\
0338$-$331&WD 0340$-$330 & 4530 (160)                        &8.00           &H &0.58           &15.70          & 64&182&7.48           &                    \\
\\
0343$-$363&WD 0345$-$362 & 4230 (100)                        &8.00           &H &0.58           &16.10          & 66&191&8.42           &                    \\
0350$-$566&WD 0351$-$564 & 3950 ( 90)                        &8.00           &H &0.58           &16.57          & 62&323&9.14           &                    \\
2211$-$392&WD 2214$-$390 & 6290 (100)                        &8.00           &H &0.59           &14.03          & 23&121&1.98           &1                   \\
2239$-$199&WD 2242$-$197 & 5400 (120)                        &8.00           &H &0.58           &14.79          & 97&162&3.61           &1                   \\
2256$-$467&WD 2259$-$465 & 4940 ( 50)                        &8.00           &He&0.57           &15.20          & 74&152&6.15           &                    \\
\\
2316$-$064&LHS 542       & 4740 ( 50)                        &8.15           &He&0.67           &15.69          & 31&250&7.29           &2                   \\
2321$-$597&WD 2324$-$595 &11180 (330)                        &8.00           &H &0.60           &11.84          & 97&272&0.45           &1                   \\
2343$-$481&WD 2346$-$478 & 4590 (120)                        &8.00           &H &0.58           &15.38          & 32& 81&7.28           &1                   \\
2346$-$550&WD 2348$-$548 & 5350 (100)                        &8.00           &H &0.58           &14.83          & 64&115&3.87           &                    \\
2349$-$031&LHS 4033      &10870 (370)                        &9.42           &H &1.34           &14.63          & 29& 97&1.75           &1, 2                \\
\\
2352$-$326&LHS 4042      & 9580 (230)                        &8.00           &H &0.60           &12.39          &100&202&0.68           &1                   \\
2354$-$211&WD 2356$-$209 & 4790 ( 50)                        &8.00           &He&0.57           &16.59          & 77&143&6.52           &4                   \\
\enddata
\tablenotetext{a}{White dwarf cooling age only, not including the main sequence lifetime.}
\tablenotetext{b}{Assumed $\logg=8$ except for stars with note (2).}
\tablecomments{
(1) \halpha\ detected spectroscopically; 
(2) $\logg$ value inferred from the trigonometric parallax; 
(3) Solution obtained with a mixed hydrogen and helium composition; 
(4) Solution obtained with $N({\rm Na})/N({\rm He})=10^{-5}$ and strong absorption feature at $V$ (see \S~4.2.1). }
\end{deluxetable}


\clearpage

\figcaption[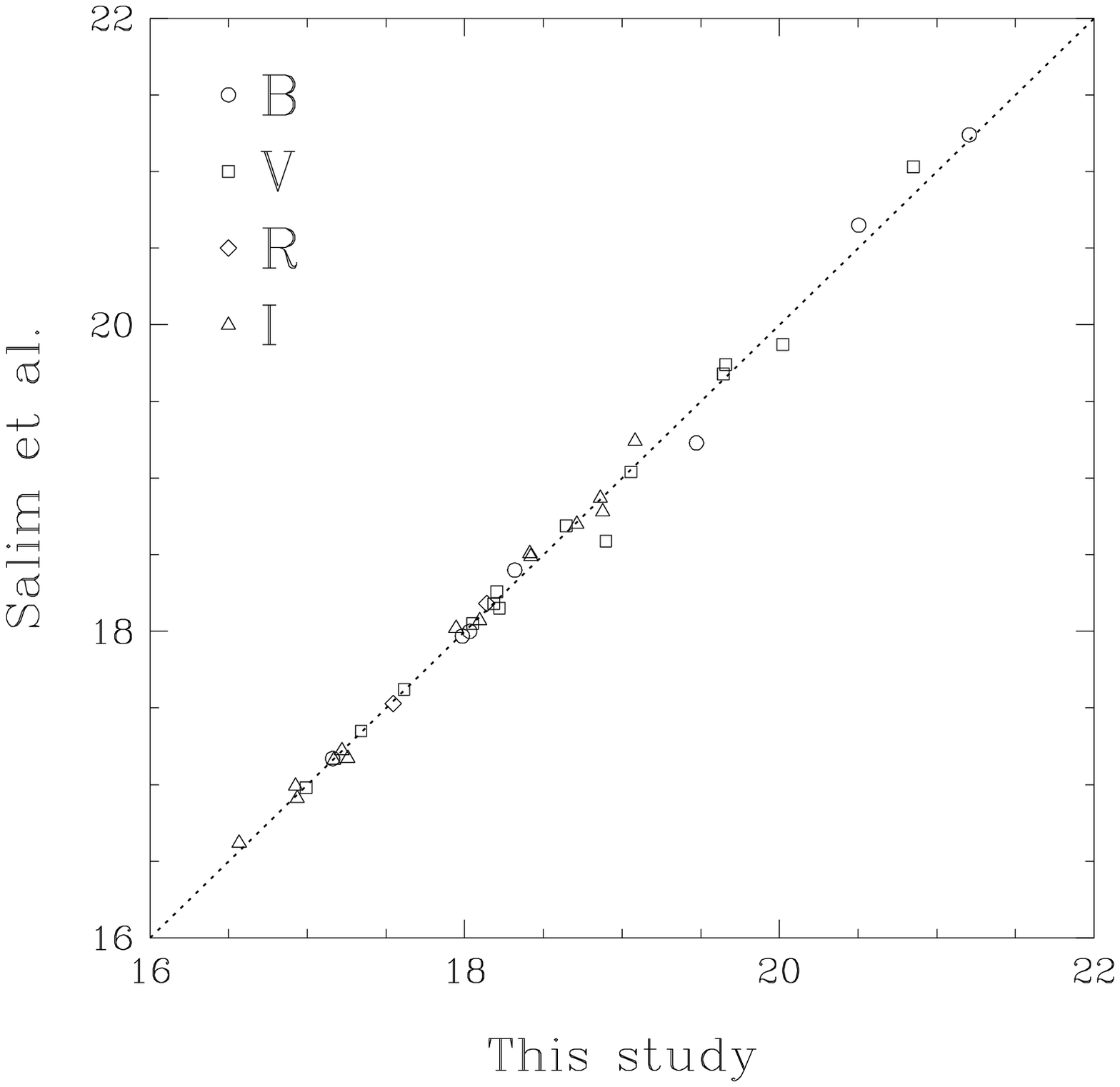] {Comparison of the optical $BVRI$ photometry
of \citet{salim04} with that obtained in this study.\label{fg:f1}}

\figcaption[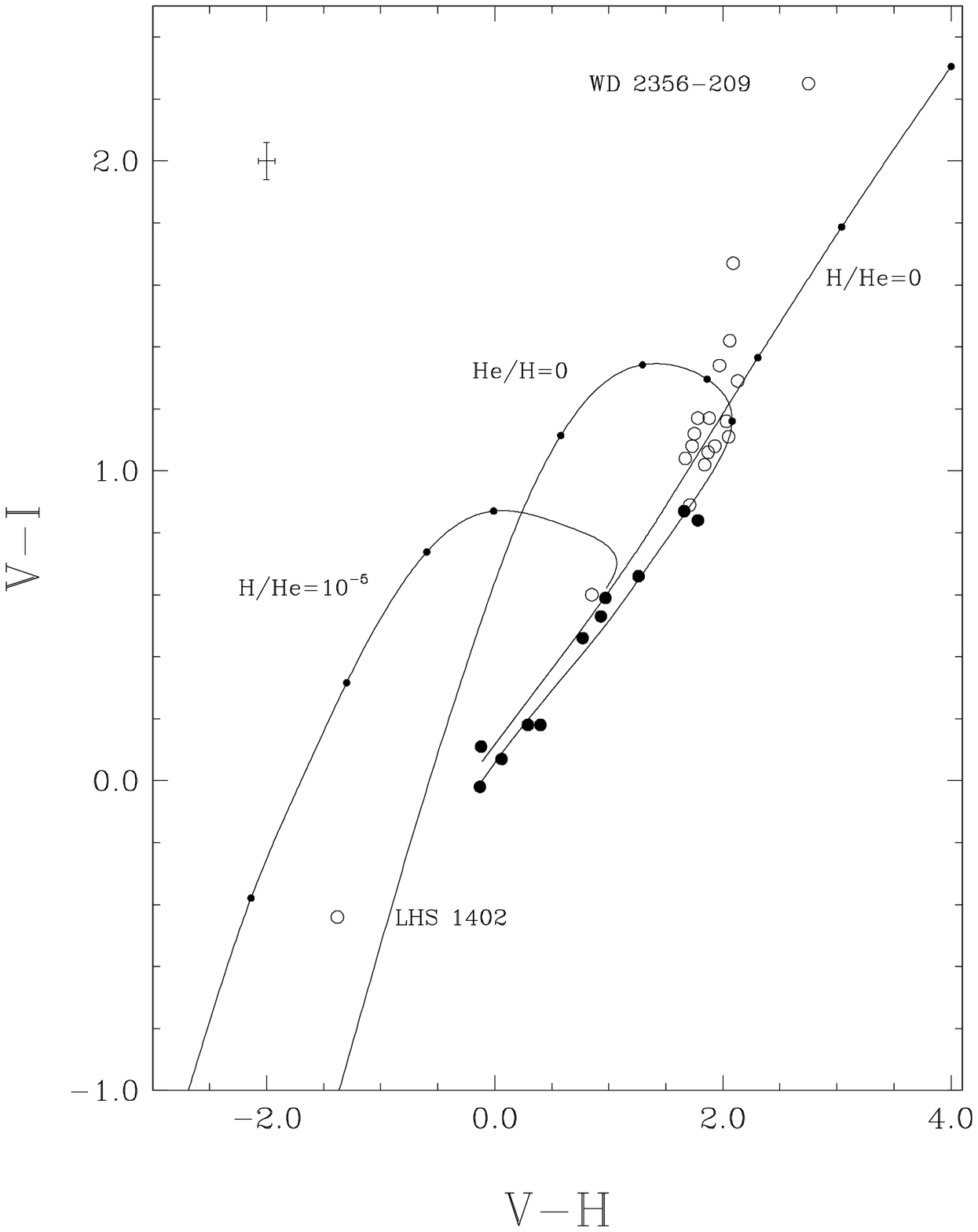] {($V$--$I$, $V$--$H$) two-color diagram for the
data set from Table 1; DA and non-DA stars are represented by filled
and open circles, respectively, and the cross indicates the size of
the average error bars. The objects marked are discussed in the
text. Theoretical colors at $\logg=8.0$ for models with pure hydrogen,
pure helium, and $\nh=10^{-5}$ atmospheric compositions are also
shown. The small dots on each sequence indicate values of $\Te$ from
3000 to 4500 K by steps of 500 K (from 3500 K only for the pure helium
sequence); the sequences start below the middle of the plot at 12,000 K.
\label{fg:f2}}

\figcaption[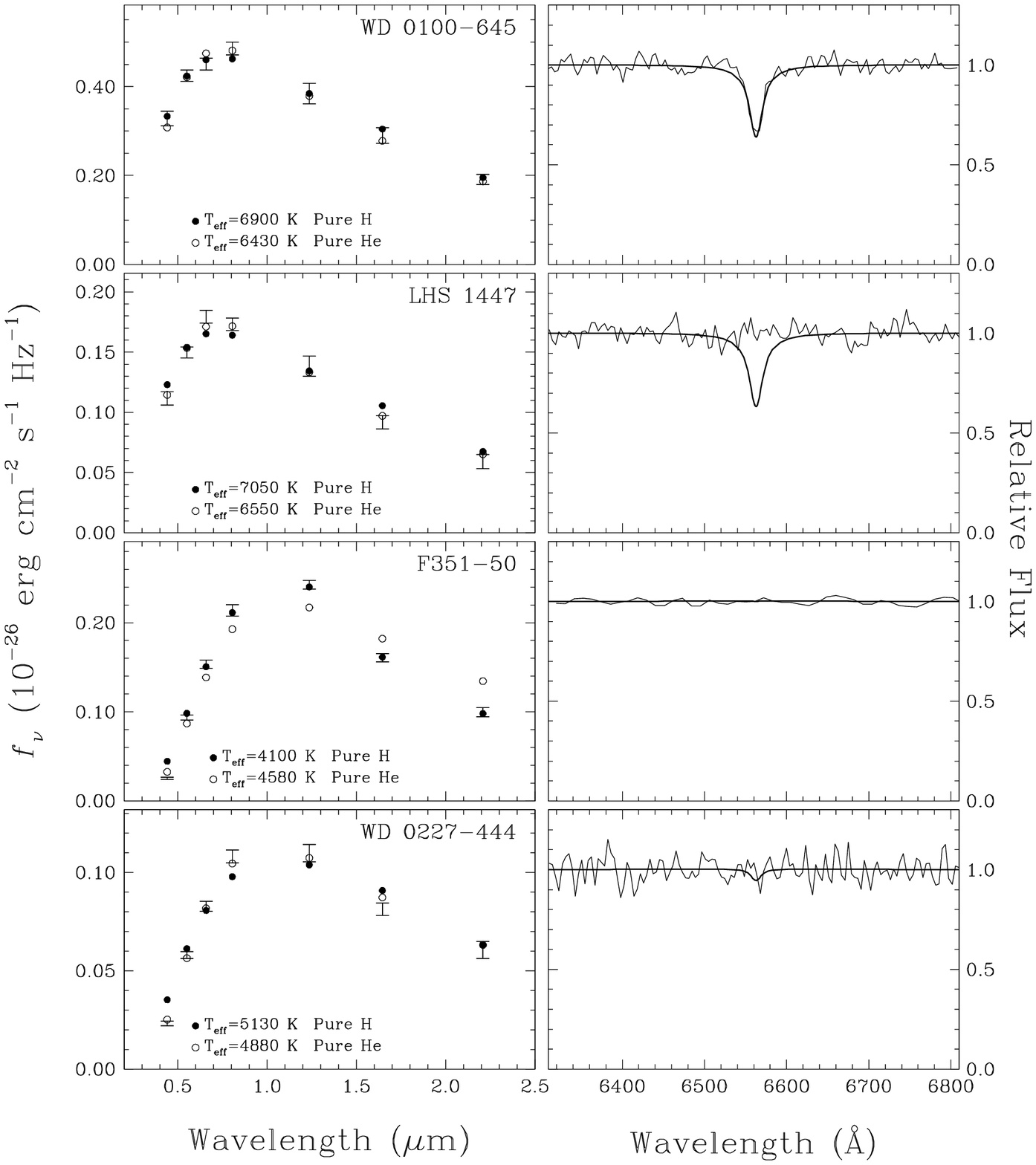] {Left panels: Sample fits to the energy distributions 
of halo white dwarf candidates with pure hydrogen models ({\it filled
circles}) and pure helium models ({\it open circles}); a value of
$\logg=8.0$ is assumed for all stars. The $BVRI$ and $JHK$ photometric
observations are represented by error bars. Right panels: Normalized
spectra near \halpha\ together with the synthetic line profiles
interpolated at the parameters obtained from the energy distribution
fits assuming a pure hydrogen atmospheric composition.
\label{fg:f3}}

\figcaption[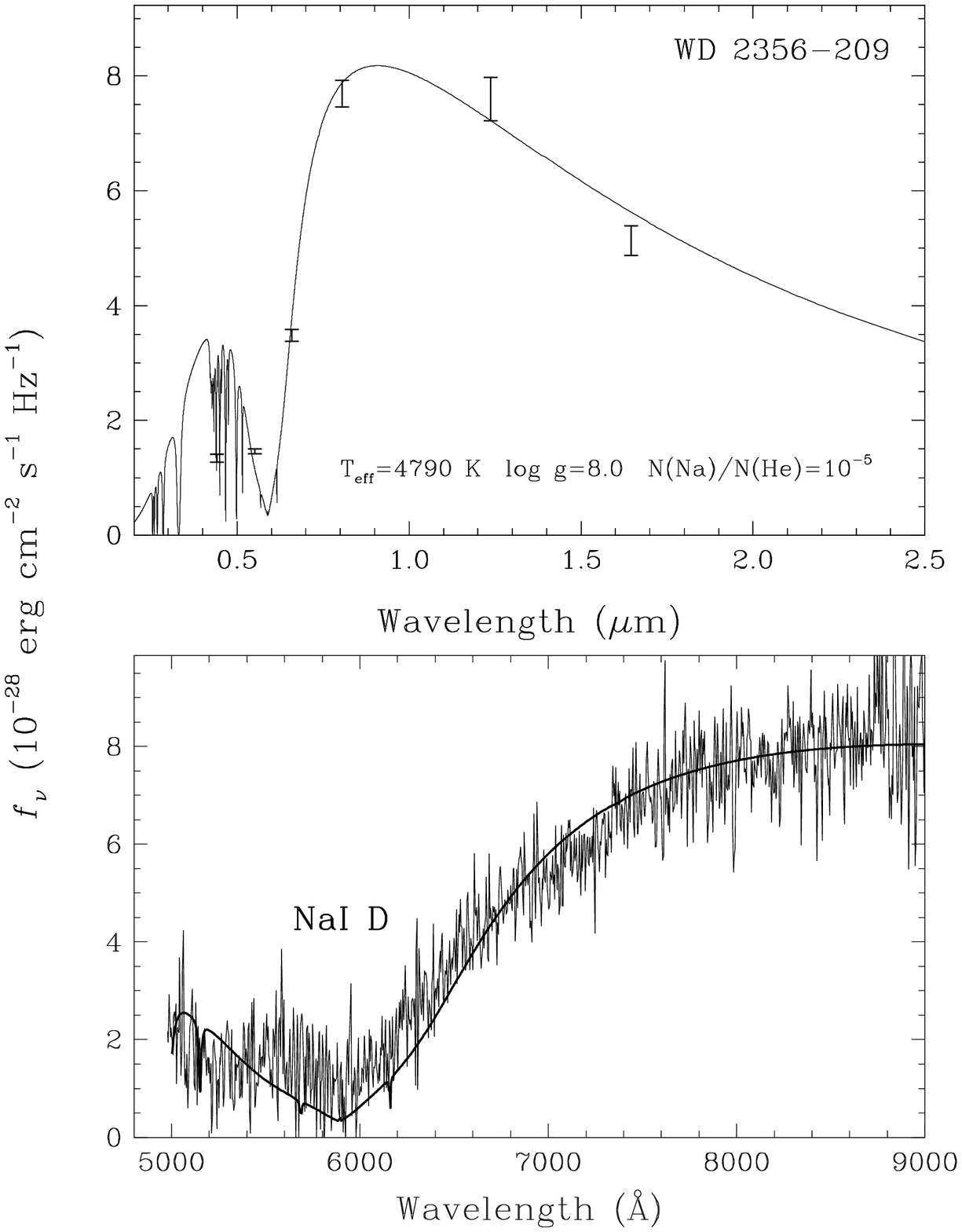] {Our best fit to the energy distribution and optical 
spectrum of WD~2356$-$209. The $BVRI$ and $JH$ photometric
observations are represented by error bars in the top panel while the
solid line corresponds to the model fluxes at the parameters indicated
in the figure; the hydrogen abundance is zero. The bottom panel shows
the observed spectrum of OHDHS together with our predicted NaI D line
profile. \label{fg:f4}}

\figcaption[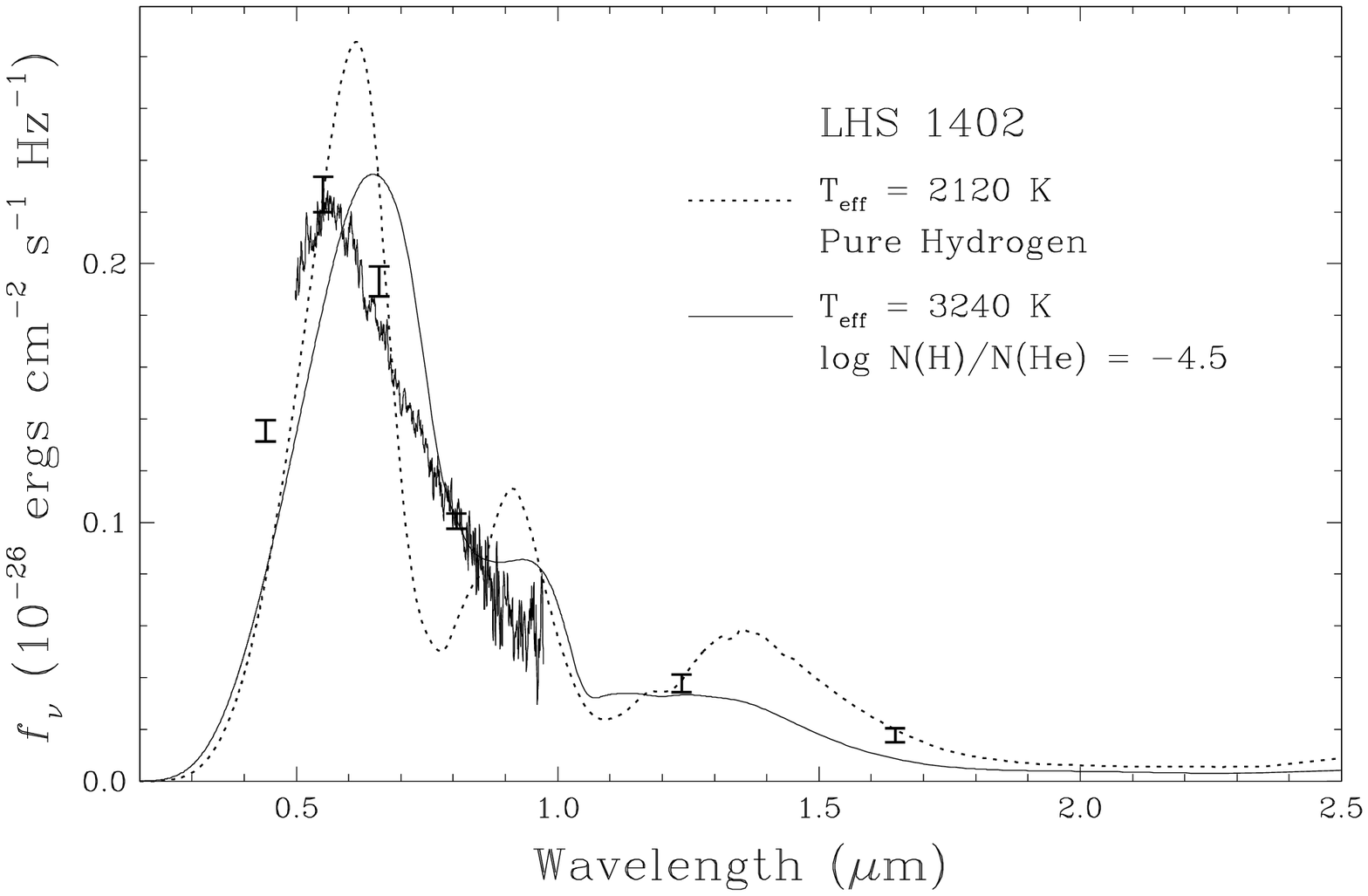] {Comparison of the best solutions for LHS 1402 
under the assumption of a pure hydrogen composition ({\it dotted
line}) and a mixed hydrogen/helium composition ({\it solid
line}). Also shown are our broadband photometry (error bars) and
optical spectrum from OHDHS. The latter suggests that LHS 1402 has a
helium-rich composition rather than a pure hydrogen
atmosphere.\label{fg:f5}}

\figcaption[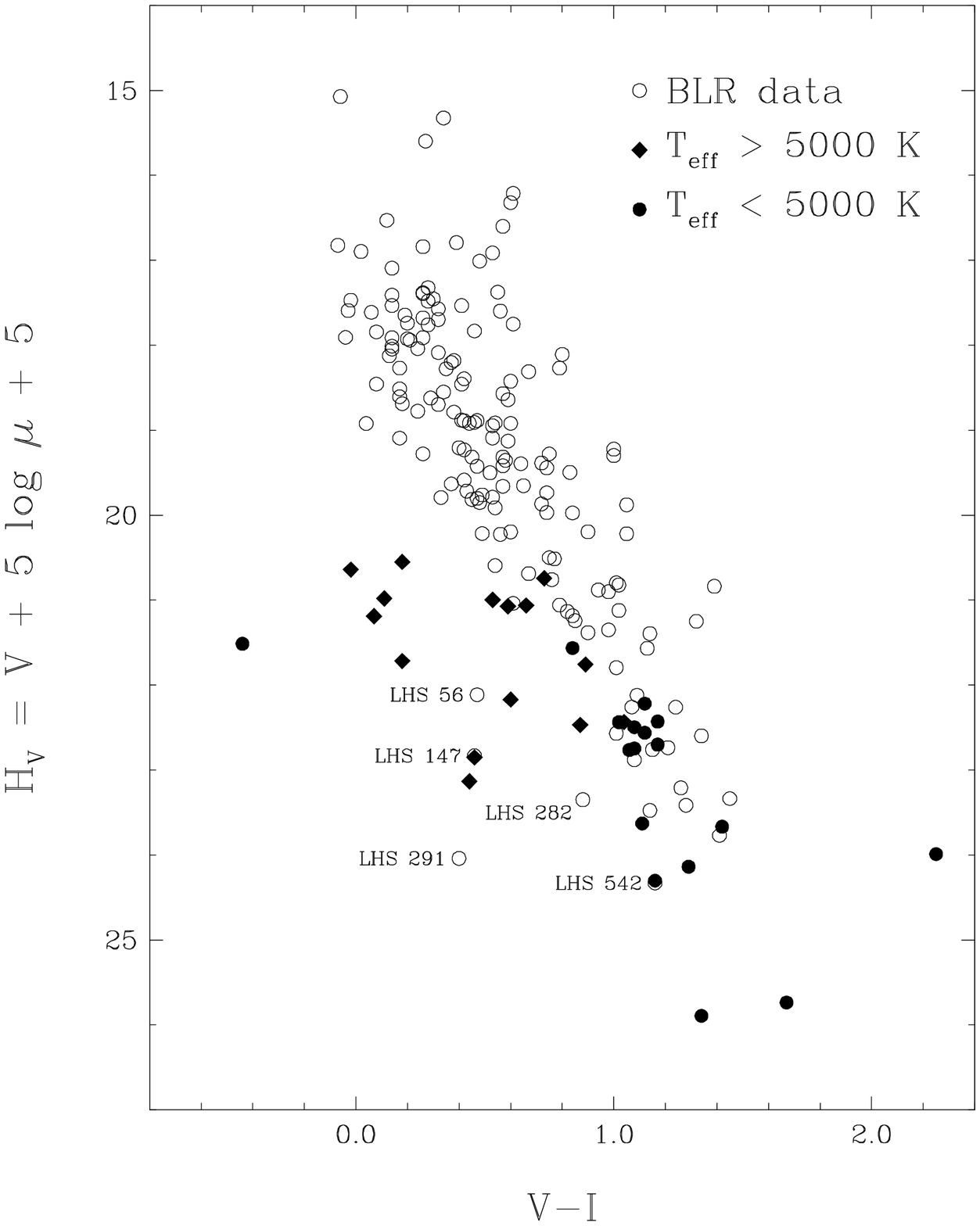] {Reduced proper motion diagram for the 
combined BRL and BLR samples ({\it open circles}) and the OHDHS sample
({\it filled symbols}). Filled circles and filled diamonds correspond
to objects below and above $\Te=5000$~K, respectively.  The objects
labeled correspond to the halo white dwarf candidates identified by
\citet{ldm89}. Note that LHS 147 and LHS 542 are in common between the
BRL and OHDHS samples. The two stars at the bottom are F351$-$50 ({\it
left}) and WD~0351$-$564 ({\it right}).\label{fg:f6}}

\figcaption[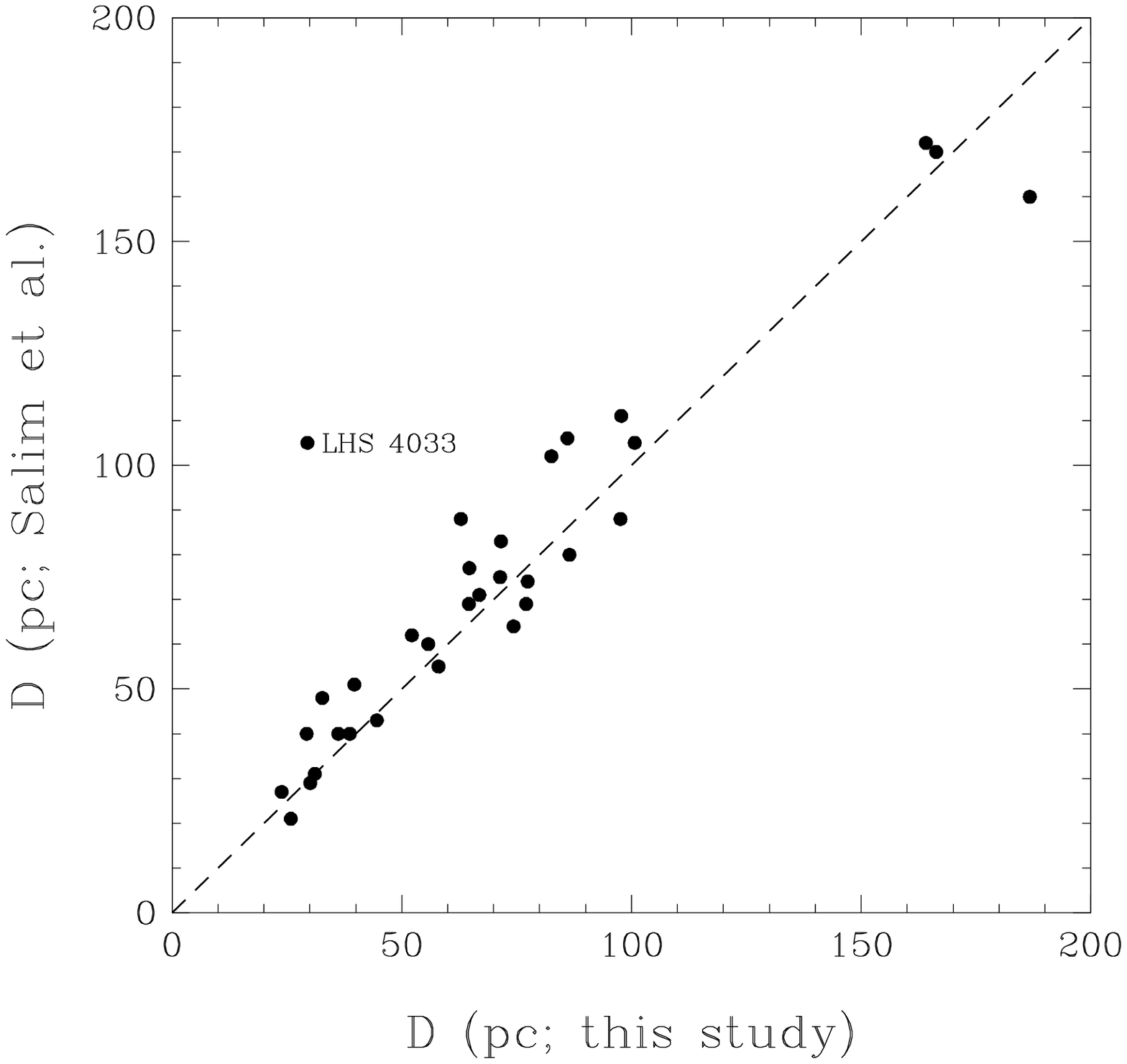] {Comparison of distances obtained
from this study with those estimated by \citet{salim04}. The discrepancy
for LHS 4033 comes from the fact that we have used the trigonometric
parallax information rather than assume a value of $\logg=8.0$.
\label{fg:f7}}

\figcaption[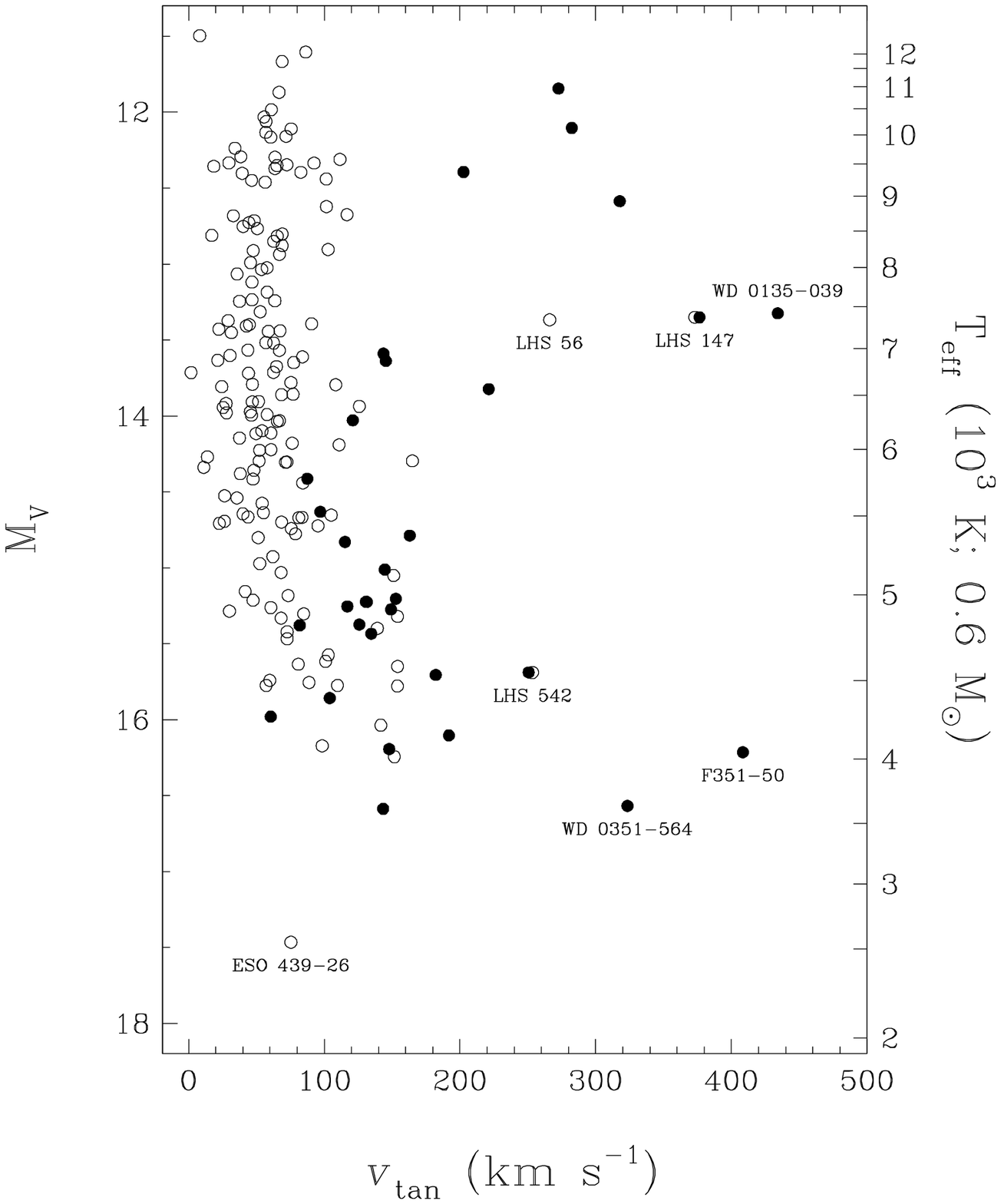] {Distribution of tangential velocities ($\vtan$) 
with the $V$ absolute magnitudes (\mv) for the white dwarfs from Table
2 ({\it filled circles}). The right axis indicates the temperature
scale for 0.6 \msun\ white dwarf models. The trigonometric parallax sample
of BLR is shown as well ({\it open circles}). The objects labeled are
discussed in the text.\label{fg:f8}}

\figcaption[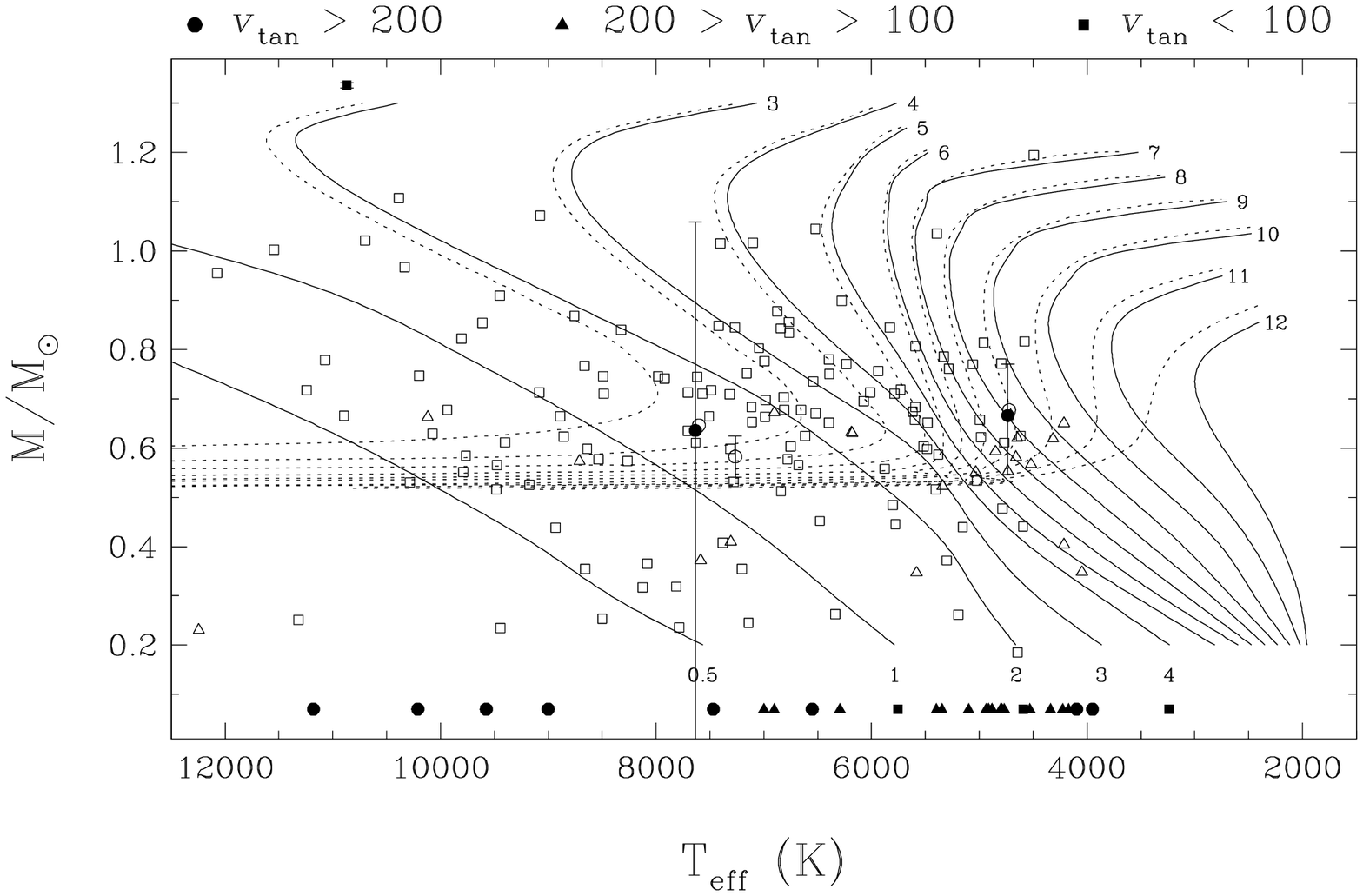] {Masses of white dwarfs in the 
trigonometric parallax sample of BLR ({\it open symbols}), and halo
white dwarf candidates from the OHDHS sample ({\it filled symbols}) as
a function of effective temperature. Various symbols explained in the
legend are used to differentiate values of the tangential velocities
(in \kms). Objects with no trigonometric parallax measurements are
shown at the bottom of the figure. Superposed are isochrones from
white dwarf cooling sequences with thick hydrogen layers ({\it solid lines});
the isochrones are labeled in units of $10^9$ years. Also shown are
the corresponding isochrones with the main sequence lifetime taken
into account ({\it dotted lines}).\label{fg:f9}}

\clearpage
\begin{figure}[p]
\plotone{f1.eps}
\begin{flushright}
Figure \ref{fg:f1}
\end{flushright}
\end{figure}

\clearpage
\begin{figure}[p]
\plotone{f2.eps}
\begin{flushright}
Figure \ref{fg:f2}
\end{flushright}
\end{figure}

\clearpage
\begin{figure}[p]
\plotone{f3.eps}
\begin{flushright}
Figure \ref{fg:f3}
\end{flushright}
\end{figure}

\clearpage
\begin{figure}[p]
\plotone{f4.eps}
\begin{flushright}
Figure \ref{fg:f4}
\end{flushright}
\end{figure}

\clearpage
\begin{figure}[p]
\plotone{f5.eps}
\begin{flushright}
Figure \ref{fg:f5}
\end{flushright}
\end{figure}

\clearpage
\begin{figure}[p]
\plotone{f6.eps}
\begin{flushright}
Figure \ref{fg:f6}
\end{flushright}
\end{figure}

\clearpage
\begin{figure}[p]
\plotone{f7.eps}
\begin{flushright}
Figure \ref{fg:f7}
\end{flushright}
\end{figure}

\clearpage
\begin{figure}[p]
\plotone{f8.eps}
\begin{flushright}
Figure \ref{fg:f8}
\end{flushright}
\end{figure}

\clearpage
\begin{figure}[p]
\plotone{f9.eps}
\begin{flushright}
Figure \ref{fg:f9}
\end{flushright}
\end{figure}

\end{document}